\documentclass[12pt,graphics]{iopart}

\newcommand{\Journal}[4]{{#1} #4 {\bf #2}, #3 }
\usepackage{iopams}  
\usepackage{graphicx}
\usepackage{color}
\usepackage{float}
\usepackage{braket}
 \expandafter\let\csname equation*\endcsname\relax
  \expandafter\let\csname endequation*\endcsname\relax
\usepackage{amsmath}
\usepackage{amssymb}
\usepackage{url}

\graphicspath{{.//}{figures//}}

\newcommand{\NIMA}{{\em Nucl. Instrum. Methods} A}

\newcommand{\NPA}{{\em Nucl. Phys.} A}
\newcommand{\PLB}{{\em Phys. Lett.}  {\bf B}}

\newcommand{\PRC}{{\em Phys. Rev.} C}

\newcommand{\PRO}{\em Phys. Rev.}

\newcommand{\EPSL}{\em Earth  Plan. Sci. Lett.}

\newcommand{\APJ}{\em ApJ}

\newcommand{\EPJA}{\em Euro. Phys. J. A}
\newcommand{\EPJC}{\em Euro. Phys. J.  C}
\newcommand{\CPC}{\em Chin. Phys.  C}

\newcommand{\JOPG}{\em Journal of Physics G}

\newcommand{\SCI}{\em Science}

\newcommand{\artn}{\mbox{$^{39}$Ar }} 
\newcommand{\arvt}{\mbox{$^{42}$Ar}}

\def\bea{\begin{eqnarray}} 
\def\eea{\end{eqnarray}} 
\newcommand{\ra}{\rightarrow }

\begin{document}

\today

\title{Spectral shapes of forbidden argon $\beta$ decays as background component for rare-event searches}

\author{J. Kostensalo, J. Suhonen and K Zuber}

\address{Department of Physics, University of Jyvaskyla,\\
P.O. Box 35, FI-40014 University of Jyvaskyla, Finland}
\ead{jouni.suhonen@phys.jyu.fi}

\address{Institut f\"ur Kern- und Teilchenphysik, Technische Universit\"at Dresden,\\
Zellescher Weg 19, 01069 Dresden, Germany}
\ead{zuber@physik.tu-dresden.de}

\begin{abstract}
The spectral shape of the electrons from the two first-forbidden unique $\beta^-$ decays of $^{39}$Ar and $^{42}$Ar 
were calculated for the first time to the next-to-leading order. Especially the spectral shape of 
the $^{39}$Ar decay can be used to characterise this background component for dark matter searches 
based on argon. Alternatively, due to the low thresholds of these experiments, the spectral shape can 
be investigated over a wide energy range with high statistics and thus allow a sensitive 
comparison with the theoretical predictions. This might lead to interesting results for the ratio of 
the weak vector and axial-vector constants in nuclei.  

\end{abstract}

\maketitle

\section{Introduction}
Liquid Argon (LAr) as a detection material is widely used in nuclear and particle physics with new applications 
coming in. The range covers calorimetry in high-energy-physics experiments at LHC all the way to 
large-scale low-background experiments for rare-event searches, especially dark matter. 
In addition, a 40 kiloton scale LArTPC detector is envisaged for the DUNE \cite{dune} long-baseline
neutrino program. The GERDA experiment \cite{gerda}, searching for neutrino-less 
double beta decay using Ge-semiconductor detectors, is cooling its detectors with LAr. 

The focus of this paper is on the low-background 
section of the potential experiments, especially dark matter experiments, of which two
ongoing LAr experiments are DEAP-3600 \cite{deap} and DarkSide-50 \cite{darkside}. 
The question arises whether there are additional  background contributions from the Ar itself.
Indeed, Ar has three long-living nuclides, $^{37}$Ar, $^{39}$Ar and $^{42}$Ar, which deserve some attention. 
First, there is the well-known $^{37}$Ar with a half-life of 35 days, which was the signal 
of the Homestake experiment for the radiochemical detection of solar neutrinos 
using a $^{37}$Cl detector \cite{cle98}. This isotope decays via electron capture (EC) without
emission of gamma rays, but it will produce X-ray lines below 10 keV. The isotope $^{37}$Ar might be produced by 
thermal neutron captures on $^{36}$Ar, but this nuclide has only a small abundance of 0.337 \%. More 
severe contaminants are the long-living nuclides \artn and \arvt, which we will discuss now in a 
bit more in detail. 

The decay of both isotopes into the ground state of the corresponding daughter nucleus
is characterised as first-forbidden unique $\beta^-$-decay: 
($\frac{7}{2}^+ \ra \frac{3}{2}^-$) for $^{39}$Ar and ($0^+ \ra {2}^-$) for $^{42}$Ar. 
The $\beta$-decay endpoints are given by  565 $\pm$ 5 keV ($^{39}$Ar) and  599 $\pm$ 6 keV ($^{42}$Ar), 
respectively \cite{ame17}. The measured half-lives for $^{39}$Ar are 
265 $\pm 30$ years \cite{Zel52} and 269$\pm$ 3 years \cite{Sto65}. Its content is 
on the level of 10 mBq/m$^3$ \cite{Loo83} and the specific activity was measured 
by the WARP experiment to be $1.01 \pm 0.02 \pm 0.08$ Bq per kg of natural Ar 
\cite{Ben07}. 
The current accepted half-life for $^{42}$Ar is 32.9 $\pm1.1$ years \cite{Sto65}. In none of 
the publications a $\beta$-spectrum of the corresponding Ar beta decay is shown.

In this paper we study the $\beta$-decay spectral shapes of the two isotopes $^{39}$Ar and
$^{42}$Ar. We calculate the shapes using up to date nuclear models. In this way we plan
to predict the expected $\beta$-spectrum shape for the Ar-based dark-matter experiments and, on the other
hand, encourage the experimentalists to provide a spectrum which can be compared with
theory.   
 
\section{Calculation of the spectral shape of \artn and \arvt}

\indent In order to simplify the nuclear $\beta^-$-decay theory enough to allow us to do practical 
calculations we use the so-called impulse approximation in which at the exact moment of decay the 
decaying nucleon only feels the weak interaction \cite{suhonen}. The strong interaction with the 
remaining $A-1$ nucleons is ignored, and thus the pion exchange and other many-body effects are 
neglected. In the impulse approximation a neutron decays into a proton via emission of a massive $W^-$ 
vector boson which in turn decays into an electron and an anti-neutrino. Due to the large mass of the 
$W^-$ boson in comparison to the energy scale of the nuclear beta decay, the $W^-$ boson couplings to 
the baryon and lepton vertices, with weak-interaction coupling strength $g_{\rm W}$, can be approximated 
as a single effective interaction vertex with effective coupling strength $G_{\rm F}$, the Fermi constant. 
When the decay process is described with the effective point-like interaction vertex, the probability of the 
electron being emitted with kinetic energy between $W_e$ and $W_e+dW_e$ is
\begin{align}
\notag
P(W_e)dW_e =& \frac{G_{\rm F}}{(\hbar c)^6}\frac{1}{2\pi^3\hbar}C(W_e) \\
& \times p_ecW_e(W_0-W_e)^2F_0(Z,W_e)dW_e \,,
\label{eq:emisprob}
\end{align}
where $p_e$ is the momentum of the electron, $Z$ is the proton number, 
$F_0(Z,W_e)$ is the Fermi function, and $W_0$ is the end-point energy of 
the $\beta$ spectrum. The nuclear-structure information is in the shape factor $C(w_e)$. 
Integrating Eq.~(\ref{eq:emisprob}) over the possible electron energies gives the total 
transition probability, and thus the half-life of the $\beta^-$ decay.

The half-life of a $\beta$ decay can be written as
\begin{equation}
t_{1/2} = \frac{\rm ln \ (2)}{\int_{m_ec^2}^{W_0} P(W_e)dW_e} := \frac{\kappa}{\tilde{C}} \,,
\label{eq:half-life}
\end{equation}
where $\tilde{C}$ is the integrated shape factor and the constant $\kappa$ has the value \cite{hardy}
\begin{equation}
\kappa = \frac{2\pi^3\hbar^7\mathrm{ln \  2}}{m_e^5c^4(G_{\rm F}
\cos \theta_{\rm C})^2}= 6147 \ \mathrm{s} \,,
\label{eq:kappa}
\end{equation}
$\theta_{\rm C}$ being the Cabibbo angle.
In order to simplify the formalism it is usual to introduce unitless kinematic 
quantities $w_e=W_e/m_ec^2$, $w_0 = W_0/m_ec^2$, and 
$p=p_ec/(m_ec^2)= \sqrt{w_e^2 -1}$. With the unitless quantities the integrated 
shape factor can be expressed as
\begin{equation}
\tilde{C} = \int^{w_0}_1 C(w_e)pw_e(w_0-w_e)^2F_0(Z,w_e)dw_e \,.
\label{eq:ctilde}
\end{equation} 
The characteristics of the electron spectrum are encoded in the shape factor $C(w_e)$, 
which can be expressed as \cite{behrens}
\begin{align}
\notag
C(w_e)=&\sum_{k_e,k_{\nu},K}\lambda_{k_e}\Big\lbrack M_K(k_e,k_{\nu})^2 +m_K(k_e,k_{\nu})^2 \\
&-\frac{2\gamma_{k_e}}{k_ew_e}M_K(k_e,k_{\nu})m_K(k_{e},k_{\nu})\Big\rbrack \,,
\label{eq:cwe}
\end{align}
where $k_{e}$ and $k_{\nu}$ (both running through 1,2,3,\dots) emerge from the partial-wave expansion 
of the electron and neutrino wave functions, $\gamma_{k_e}=\sqrt{k_e^2-(\alpha Z)^2}$, and 
$\lambda_{k_e}=F_{k_e-1}(Z,w_e)/F_0(Z,w_e)$ is the Coulomb function where $F_{k_e-1}(Z,w_e)$ is the 
generalized Fermi function. The largest contributions to the sum of Eq.~(\ref{eq:cwe}) come from 
the terms with minimal angular-momentum transfer, so in the case of the unique forbidden decays 
studied in this work, we only consider the sum that satisfies the condition $k_e+k_{\nu}=K+2$. The 
quantities $M_K(k_e,k_{\nu})$ and $m_K(k_e,k_{\nu})$ have lengthy expressions involving kinematic and 
nuclear form factors. The explicit expressions can be found from \cite{behrens} (the expressions were 
also given in the recent article \cite{Haaranen2017}). The form factors appearing in these expressions 
can be expanded as power series of $qR/\hbar$ as
\begin{equation}
\mathcal{F}_{\rm KLS}(q^2)=\sum_N \frac{(-1)^N(2L+1)!!}{(2N)!!(2L+2L+1)!!}(qR/\hbar)^{2N}\mathcal{F}^{(N)}_{\rm KLS} \,,
\label{eq:fex}
\end{equation}
where $q=|p_e+p_{\nu}|$ and $R$ is the nuclear radius. In practical calculations the quantities 
$M_K(k_e,k_{\nu})$ and $m_K(k_e,k_{\nu})$ are expanded as a power series of the quantities 
$\eta_{1,2,3,4,5}=\alpha Z$, $p_eR/\hbar$, $qR/\hbar$, $m_ecR/\hbar$, and $W_eR/\hbar c$. $M_K(k_e,k_{\nu})$ 
and $m_K(k_e,k_{\nu})$ consist of terms proportional to $\prod_i \eta_i^{\alpha_i} \mathcal{F}_{\rm KLS}$, 
where $\alpha_i=0,1,2,3,4,5$. In the case of the unique decays the often used leading-order approximation 
takes into account only the $(p_eR/\hbar)^{k_e-1}(qR/\hbar)^{k_{\nu-1}}$ term, while the here adopted 
next-to-leading-order treatment \cite{Haaranen2017} takes into account also the 
$(p_eR/\hbar)^{k_e-1}(qR/\hbar)^{k_{\nu-1}}\eta_j$ terms. The leading-order and 
next-to-leading-order expressions of the $\beta$-decay shape factor are discussed in detail 
in Ref. \cite{Haaranen2017} for a more general framework including also
the (more involved) non-unique forbidden decays.

In the impulse approximation the nuclear form factors can be replaced by nuclear matrix elements (NMEs) 
\cite{behrens}. In the leading-order approximation there is only one NME contributing to the shape 
factor for the unique decays. Therefore, the spectrum-shape and the half-life are proportional to 
$g_{\rm A}^{-2}$, where $g_{\rm A}$ is the weak axial-vector coupling constant. 
When the next-to-leading-order terms are taken into account, the number of contributing 
NMEs increases to five, with each NME carrying a prefactor $g_{\rm A}$ or $g_{\rm V}$ (weak vector coupling
constant). As a result, the shape factor can be expressed as a decomposition 
\begin{equation}
C(w_e)= g_{\rm V}^2C_{\rm V}(w_e)+g_{\rm A}^2C_{\rm A}(w_e)+
g_{\rm V}g_{\rm A}C_{\rm VA}(w_e) \,.
\label{eq:decompc}
\end{equation}
In the here adopted next-to-leading-order theory the dependence on the weak coupling constants of the 
spectrum shape for the studied decays of Argon isotopes is, at least theoretically, non-trivial.

		\begin{figure*}
	\centering	
	\includegraphics[width=0.8\textwidth]{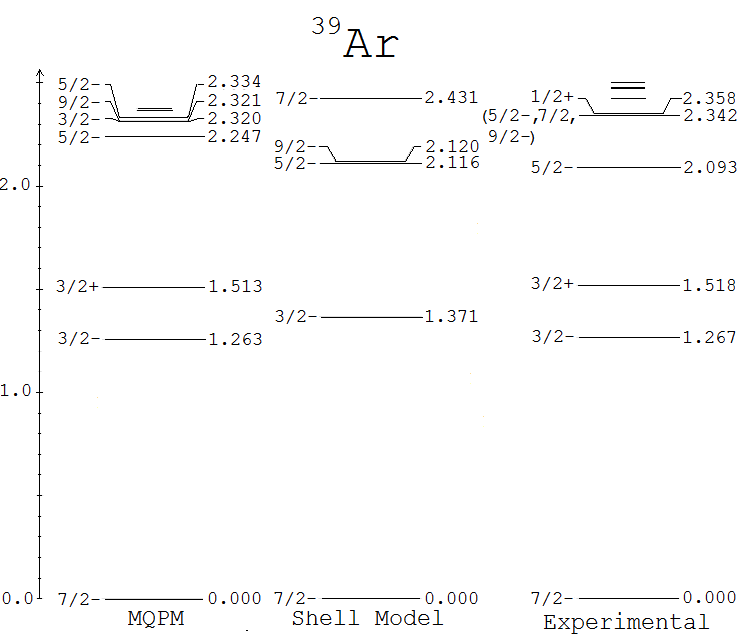}
	\caption{ The excitation spectrum of $^{39}\rm Ar$ calculated using the MQPM and the shell model. 
The experimental excitation spectrum is from \cite{nndc}.
	\label{fig:spec}}
	\end{figure*}

The choice of a nuclear model enters the picture when calculating the one-body transition densities (OBTDs) 
needed for the evaluation of the NMEs related to the transition. In this work the wave functions of the 
initial and final states were calculated using the microscopic quasiparticle-phonon model (MQPM) 
\cite{mqpm1,mqpm2} and the nuclear shell model. The MQPM is a fully microscopic model which can be used 
to describe spherical odd-$A$ nuclei, in the case of this work $^{39}\rm Ar$ and $^{39}\rm K$. In the MQPM 
the states of the odd-$A$ nucleus are built from BCS quasiparticles and their couplings to QRPA phonons. 
The quasiparticles and phonons emerge from the calculations done on the neighboring even-even reference 
nucleus (here $^{38}\rm Ar$). The MQPM has previously been used in the calculations of forbidden 
beta-decay spectrum shapes in Refs. \cite{Haaranen2017,nonunique,Kostensalo2}. 

The practical application of the MQPM model follows the same basic steps as in the earlier studies 
(see Refs.~\cite{Haaranen2017,mqpm2, nonunique,Kostensalo2,mustonen06, haaranen13}) including the use 
of the Bonn one-boson-exchange potential with G-matrix techniques \cite{mqpm2}. The single-particle 
energies, used to solve the BCS equations, 
were calculated using the Coulomb-corrected Woods-Saxon potential with the Bohr-Mottelson 
parametrization \cite{bohr1969}. The valence space spanned the orbitals $0s0p1s0d0f1p0g_{9/2}$. The BCS 
one-quasiparticle spectra were tuned by adjusting manually some of the key single-particle energies 
to get a closer match between the low-lying one-quasiparticle states and the corresponding experimental ones. 
The empirical pairing gaps, computed using the data from \cite{ame12}, were 
adjusted to fit the computed ones by tuning the pairing strength parameters $g_{\rm pair}^{\rm p}$ and 
$g_{\rm pair}^{\rm n}$ for protons and neutrons separately. In the MQPM calculations an extended cutoff 
energy of 6.0 MeV was used for the QRPA phonons instead of the 3.0-MeV cutoff energy used in Refs. 
\cite{nonunique,Kostensalo2}.

The shell-model OBTDs were calculated using the shell-model code NuShellX@MSU \cite{nushellx} with the 
effective interaction sdpfnow \cite{sdpfnow}, tuned for the $1s0d0f1p$ valence space. Since performing a 
shell-model calculation in the entire half-filled $1s0d0f1p$ valence space would be impossible due to 
the extremely large dimensions of the shell-model Hamiltonian matrix, some controlled truncations 
were made. We limited the protons to the $sd$-shell, leaving the $fp$-shell empty, and forced a complete filling 
of the $sd$-shell for neutrons, leaving the entire $fp$-shell as the neutron valence space. These truncations are 
reasonable since we are only interested in the ground-state wave functions, to which configurations 
opening the neutron $sd$-shell and proton configurations with protons in the $fp$-shell have very small 
contributions due to the considerable energy gap between the two shells. The adopted shell-model Hamiltonian 
predicts correctly the spin-parities of the low-lying states in the studied nuclei. As an example, the energy 
spectra of $^{39}\rm Ar$, predicted by the MQPM and shell model, are compared 
with the experimental spectrum \cite{nndc} in Fig. \ref{fig:spec}. In the shell-model 
spectrum the positive-parity states are missing, since the odd neutron is restricted to the $fp$-shell.

		\begin{figure}
	\centering	
	\includegraphics[width=0.6\textwidth]{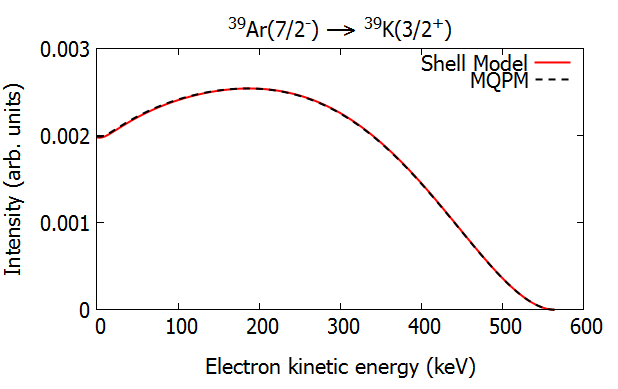}
	\caption{Electron spectrum of the ground-state-to-ground-state decay of $^{39}\rm Ar$ 
calculated using the MQPM and the shell model with the effective interaction sdpfnow \cite{sdpfnow}. 
The experimental spectrum is from NNDC \cite{nndc}.
	\label{fig:el39}}
	\end{figure}

			\begin{figure}
	\centering	
	\includegraphics[width=0.6\textwidth]{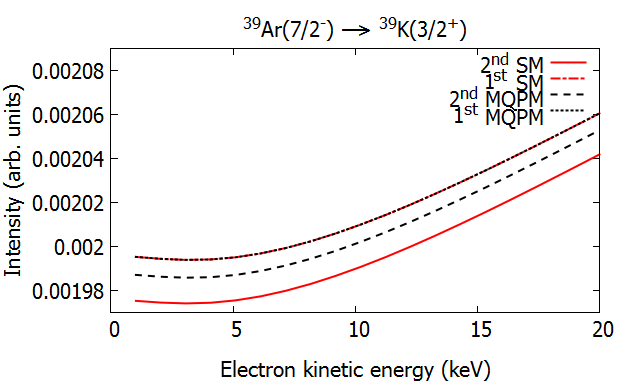}
	\caption{Zoom-in of the low-energy end of the electron spectrum of $^{39}\rm Ar$ calculated 
using the shell model (SM) and the microscopic quasiparticle-phonon model (MQPM) with leading-order (1st) 
and next-to-leading-order (2nd) terms of the shape factor $C(w_e)$ included. The values 
$g_{\rm A}=g_{\rm V}=1.00$ were adopted. Note the range on the $y$-axis. \label{fig:order} }
	\end{figure}
	
				\begin{figure}
	\centering	
	\includegraphics[width=0.6\textwidth]{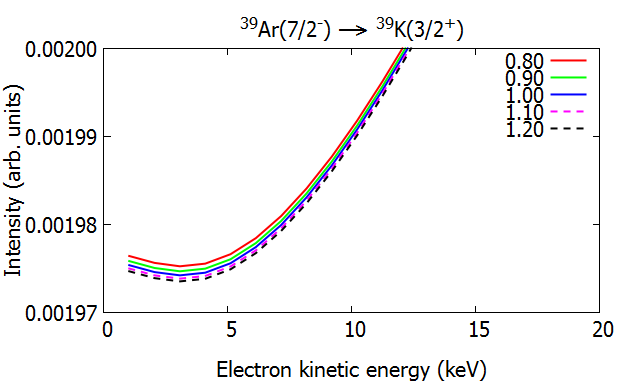}
	\caption{Zoom-in of the low-energy end of the electron spectrum of $^{39}\rm Ar$ calculated 
using the shell model. The color/dash coding refers to the ratio  $g_{\rm A}/g_{\rm V}$. Note 
the range on the $y$-axis. \label{fig:ga} }
	\end{figure}
	
					\begin{figure}
	\centering	
	\includegraphics[width=0.6\textwidth]{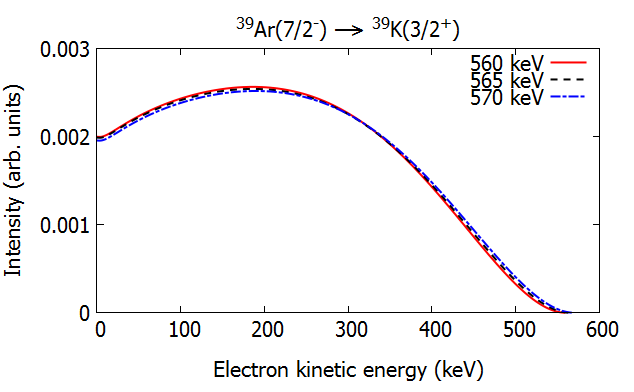}
	\caption{Electron spectrum of $^{39}\rm Ar$ calculated using the shell model and the 
experimental Q-value $656\pm 5$ keV \cite{ame12}. The next-to-leading-order terms are included 
and the values $g_{\rm A}=g_{\rm V}=1.00$ were adopted. \label{fig:q}}
	\end{figure}
	
		\begin{figure}
	\centering	
	\includegraphics[width=0.6\textwidth]{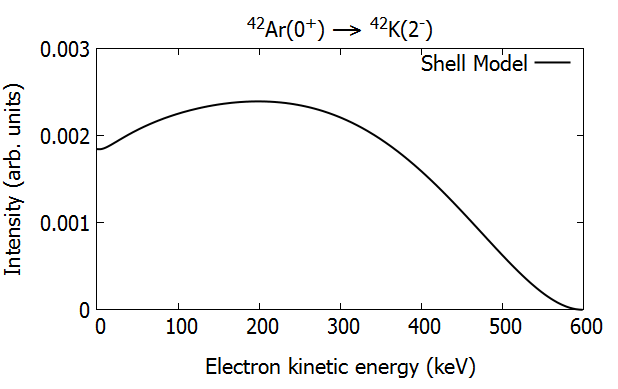}
	\caption{Electron spectrum of the ground-state-to-ground-state decay of $^{42}\rm Ar$ calculated 
using the shell model with the effective interaction sdpfnow \cite{sdpfnow}. The experimental 
spectrum is from NNDC \cite{nndc}.
	\label{fig:el42}}
	\end{figure}
	
	The electron spectrum of the $^{39}\rm Ar$ decay is presented in Fig.~\ref{fig:el39}. Here the 
next-to-leading-order terms are taken into account and the values $g_{\rm V}=g_{\rm A}=1$ were adopted for 
the weak coupling constants. The spectra were calculated also using the $g_{\rm A}/g_{\rm V}$ ratios $0.8-1.2$ 
but the spectra coincided perfectly, meaning that the possible dependence of the spectrum shape on the weak 
coupling constants is very weak. The spectra calculated using the shell model and the MQPM agree also 
extremely well, suggesting that the uncertainty related to the wave functions of the initial and final states 
is tiny. The difference between the two spectra is largest at the low-energy end, while at the high-energy 
end the spectra coincide perfectly. In Fig.~\ref{fig:order} a zoom-in of the low-energy part of the spectrum 
is presented, with the next-to-leading-order terms of the shape factor either included or neglected. The 
difference in the intensities predicted by the two models, with the next-to-leading-order terms included, is 
at most $0.5\%$ of the intensity. 
	
	When the next-to-leading-order-terms are taken into account, the dependence of the shape-factor 
on the ratio $g_{\rm A}/g_{\rm V}$ becomes non-trivial for the unique forbidden decays. The difference between 
the $^{39}\rm Ar$ spectra calculated using different values of $g_{\rm A}/g_{\rm V}$ is largest at the 
low-energy end. The magnified low-energy shell-model electron spectrum is presented in Fig.~\ref{fig:ga}. 
With larger quenching of $g_{\rm A}$, the intensity of the low-energy electrons increase slightly, but the 
difference is negligible within the accuracy of practical measurements. 
	
	The final source of uncertainty of the spectrum shape is the uncertainty related to the Q-value. 
The current experimental Q-value of the $^{39}\rm Ar$ decay is $565\pm 5$ keV. The shell-model spectra with 
Q-values 560, 565, and 570 keV are plotted in Fig.~\ref{fig:q}. Here a small difference between the 
spectra can be seen without further magnification. The uncertainty related to the Q-value seems to be by 
far the largest contributor to the uncertainty in the theoretical spectrum-shape. The change in Q-value 
stretches the spectrum, but the overall shape does not change.
	
	For the $^{42}\rm Ar$ ground-state-to-ground-state decay the MQPM cannot be applied. The 
uncertainties related to the wave functions were negligible in the case of $^{39}\rm Ar$, as well as in the 
several other first-forbidden unique decays studied in Ref.~\cite{Kostensalo2}, so the application of only 
one nuclear model should be sufficient. The electron spectrum is presented in Fig.~\ref{fig:el42}. Again, 
there is no significant dependence on the values of $g_{\rm A}$ and $g_{\rm V}$. For the decay of the 
$^{42}\rm Ar$ isotope the experimental uncertainty of the Q-value is at 6 keV, similar to that of the 
$^{39}\rm Ar$ isotope, and thus the uncertainty of the spectrum-shape is also of a similar magnitude.
	
\section{Summary and conclusions}
In this paper the $\beta$ spectra of first-forbidden unique decays of $^{39}$Ar and $^{42}$Ar were 
calculated for the first time using a next-to-leading-order weak theory. The involved nuclear
wave functions were computed by using the nuclear shell model and the microscopic quasiparticle-phonon
model. The major uncertainty in studies of the speactral shapes is related to the uncertainty
of the Q-value. Among the different calculations only very tiny differences could
be observed at low energy of the $\beta$ spectrum, which can be studied in detail with running or 
future LAr dark-matter detectors, as they are sensitive in this energy range. 
This high sensitivity might allow to explore the $g_{\rm A}/g_{\rm V}$ ratio of the weak coupling
constants. The higher-energy part of the $\beta$ spectrum of $^{39}$Ar can already be seen in the released
spectra of the GERDA experiment, but the lower-energy part might be disturbed due to dead-layer effects 
of the Ge detectors. To summarise, the calculated $\beta$ spectra can help to characterise 
the background contribution of the two studied isotopes for dark-matter searches based on LAr 
dark-matter detectors. On the other hand, these experiments can investigate the presently predicted 
spectral shape in detail as the expected decay rate, especially for $^{39}$Ar, is reasonably high
in these detectors. 

\section{Acknowledgement}
This work was partly supported by the Academy of Finland under the Finnish Center of Excellence 
Program 2012-2017 (Nuclear and Accelerator Based Program at JYFL).

\end{document}